\documentclass{ws-ijb}
\usepackage{tabulary}
\usepackage{multirow}
\usepackage{float}

\begin{document}
\title{Predicting the stability of profiling signals of small RNAs}
% \author{Qiuyun Li\thanks{Rose-Hulman Institute of Technology, Terre Haute, IN 47803}, Manda Riehl\footnotemark[1]}
\author{QIUYUN LI}

\address{Mathematics Department, Rose-Hulman Institute of Technology, 5500 Wabash Avenue\\
Terre Haute, IN 47803,USA%\footnote{State completely without 
%abbreviations, the affiliation and mailing address, including country. 
%Typeset in 8 pt italic.}\\ 
%\email{liq1@rose-hulman.edu%\footnote{Typeset author's e-mail 
%address in 8pt italic.}} 
}

\author{MANDA RIEHL}

\address{Mathematics Department, Rose-Hulman Institute of Technology, 5500 Wabash Avenue\\
Terre Haute, IN 47803,USA\\
riehl@rose-hulman.edu }

\maketitle

\begin{abstract}
{\bfseries Abstract.}\quad
Profiling is a process that finds similarities between different RNA secondary structures by extracting signals from the Boltzmann sampling. The reproducibility of profiling can be identified by the standard deviation of number of features among Boltzmann samples. We found a strong relationship between the frequency of each helix class and its standard deviation of the frequency upon repeated Boltzmann sampling. We developed a perturbation technique to predict the stability of these featured helix classes without the need for repeated Boltzmann sampling, with accuracy between 84\% and 94\%, depending on the type of RNA.  Our technique only requires 0.2\% of the computation time compared to one profiling process.
\end{abstract}
\section{Introduction}

RNA is an important biological macromolecule and is known as an intermediate messenger between DNA and protein, but many small RNAs have been found to have catalytic properties for biochemical reactions like enzymes and also function in regulating gene expression. For example, tRNAs bind to amino acids and read the gene code from mRNAs, THF riboswitches bind to tetrahydrofolate\cite{ribo1}, TPP riboswitches bind to thiamine pyrophosphate\cite{ribo} and Qrrs serve as a center in the quorum sensing regulatory circuit (\cite{qrr}, \cite{qrr2}, \cite{qrr3}). The secondary structure is important for the RNA’s non-coding function. 

Different structures require different energies in the bonds of the RNA. Structures with lower free energy will be more stable. GTfold is software that samples structures from the Boltzmann ensemble based on the energy of the possible structures.\cite{gtfold} RNAStructProfiling is software that uses GTfold to sample 1000 secondary structures, and processes the structures into helix classes and profiles. It extracts useful structural signals from the noisy Boltzmann sample.\cite{profile} 

\begin{figure}
\centering
 \includegraphics[width=10cm]{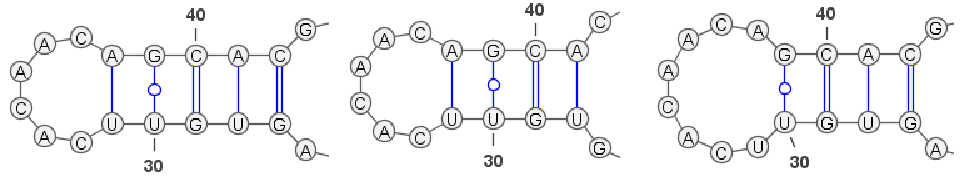}
 \caption{Helix (27, 42, 5), (28, 41, 4) and (27, 42, 4).}
 \label{helix}
\end{figure}

In a single strand of RNA, base pair bonds can form between AG, GC and GU. Base pair bonds can be represented by the indices of its two bases. A helix is a consecutive series of base pairs, and can be represented by their starting index, ending index and number of base pairs. For example, helices (27, 42, 5) is the collection of base pair (27,42), (28,41), (29,40), (30,39), and (31,38). When comparing different structures in a Boltzmann sample, a helix in one structure may be a subset of a helix in another structure. For example, in Fig. \ref{helix}, helix (27, 42, 5), (28, 41, 4), (27, 42, 4) are subsets of helix (27, 42, 5). In this case, (27, 42, 5) is a \emph{helix class}. Helix classes with high frequencies in the sample are selected. The selection of cutoff is found by calculating Shannon entropy, which represents how much information the helix classes give.
The helix classes from 1000 structures are sorted by frequency in descending order. The probability $p_k$ that the $k^{th}$ helix occurs is normalized by the highest frequency, so
		\begin{equation}\label{pk}
        p_{k}=\frac{f_k}{f_1}. 
        \end{equation} 
        The Shannon entropy of the $k^{th}$ class can be calculated using the equation
        \begin{equation}\label{Hk}
         H_{k}=-p_{k}ln(p_{k})-(1-p_{k})ln(1-p_{k}).
        \end{equation}
        The cumulative average Shannon entropy at $k^{th}$  helix class can be calculated using the equation 
        \begin{equation}\label{hk}
        h_{k}=\frac{1}{k}\sum_{i=1}^k H_{i}.
        \end{equation}
In order to make the selected $k$ helix classes give us maximum information, the threshold is set at the point where the cumulative average Shannon entropy, $h_k$, reaches its maximum. The selected helix classes are called \emph{features}, and then reported to users.

If a software user repeated the process with the same RNA sequence, they may not expect that the result could be different. When they run the program again, the software generate a new Boltzmann sample, which will differ across replications. The reproducibility of RNAStructProfiling has been measured by taking the standard deviation of the number of features across multiple Boltzmann samples.

We developed an algorithm to predict the reproducibility of the number of features without taking different samples. This extra feature could be added to the software and will give users an expectation about the stability ahead of time. Generating the 1000-structure sample takes most of the run time in the total process. Our prediction algorithm avoids repeated sampling and gives a reliable prediction; on average, the prediction of stability only takes 0.2\% of the time used for one profiling process.

\section{Stability}

In the previous research, Rogers and Heitsch measured reproducbility for 15 different RNA sequences with 25 replications each. They showed that profiling is reproducible by calculating the standard deviation of the number of features and selected profiles. Their standard deviations range from 0 to 0.8 in the 15 RNA sequences. Their results showed that only one of the RNA, THF in Streptococcus uberis, has 0 standard deviation for the number of features.\cite{profile} We tested stability on a wider range of RNAs: 233 Qrr RNAs, 130 THF RNAs, 124 tRNAs, and 177 TPP RNAs\cite{data}. 

\begin{figure}
    \centering
    \includegraphics[width=14.5cm]{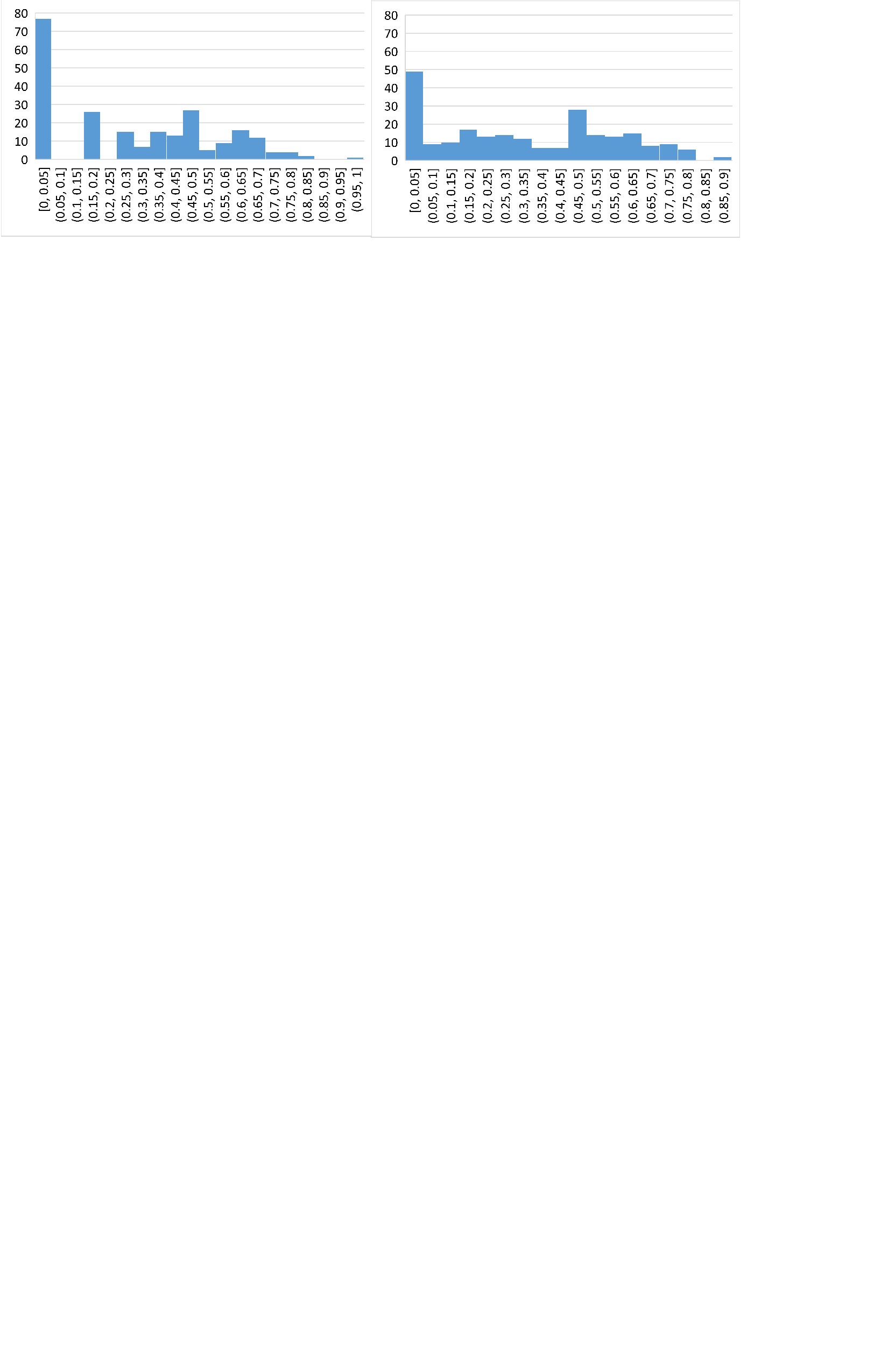} 
    \vspace{-19cm}
    \caption{Histogram for the standard deviations of number of features for 233 Qrr RNAs with 25 (left) and 100 (right) replications.}
    \label{histogram}
\end{figure}

In our investigation on the stability of 233 Qrr RNAs, a large number of the RNAs show extremely high stability. We tested them with different number of replications, and the result is shown in Fig. \ref{histogram} and Table \ref{stablility}. In fact, 33 of them have their number of features constant even with 500 replications. We chose to measure stability with the standard deviation of the number of the number of features for 100 replications. If the standard deviation of the number of features for 100 replication is 0, the result is identified as \textit{stable}. Otherwise, it is \textit{unstable}.

\begin{table}[]
\centering
\caption{Number of Qrr RNAs with standard deviation of 0 on number of features, with 25, 100, or 500 replications out of all 233 Qrr RNAs.}
\begin{tabulary}{10cm}{|C|C|}
\hline
 Number of replications & Number of Qrr RNAs that have 0 standard deviation for number of features \\ \hline
 25&77  \\ \hline
 100&49  \\ \hline
 500&33  \\ \hline
\end{tabulary}
\label{stablility}
\end{table}

\begin{figure}
    \centering
    \includegraphics[width=12cm]{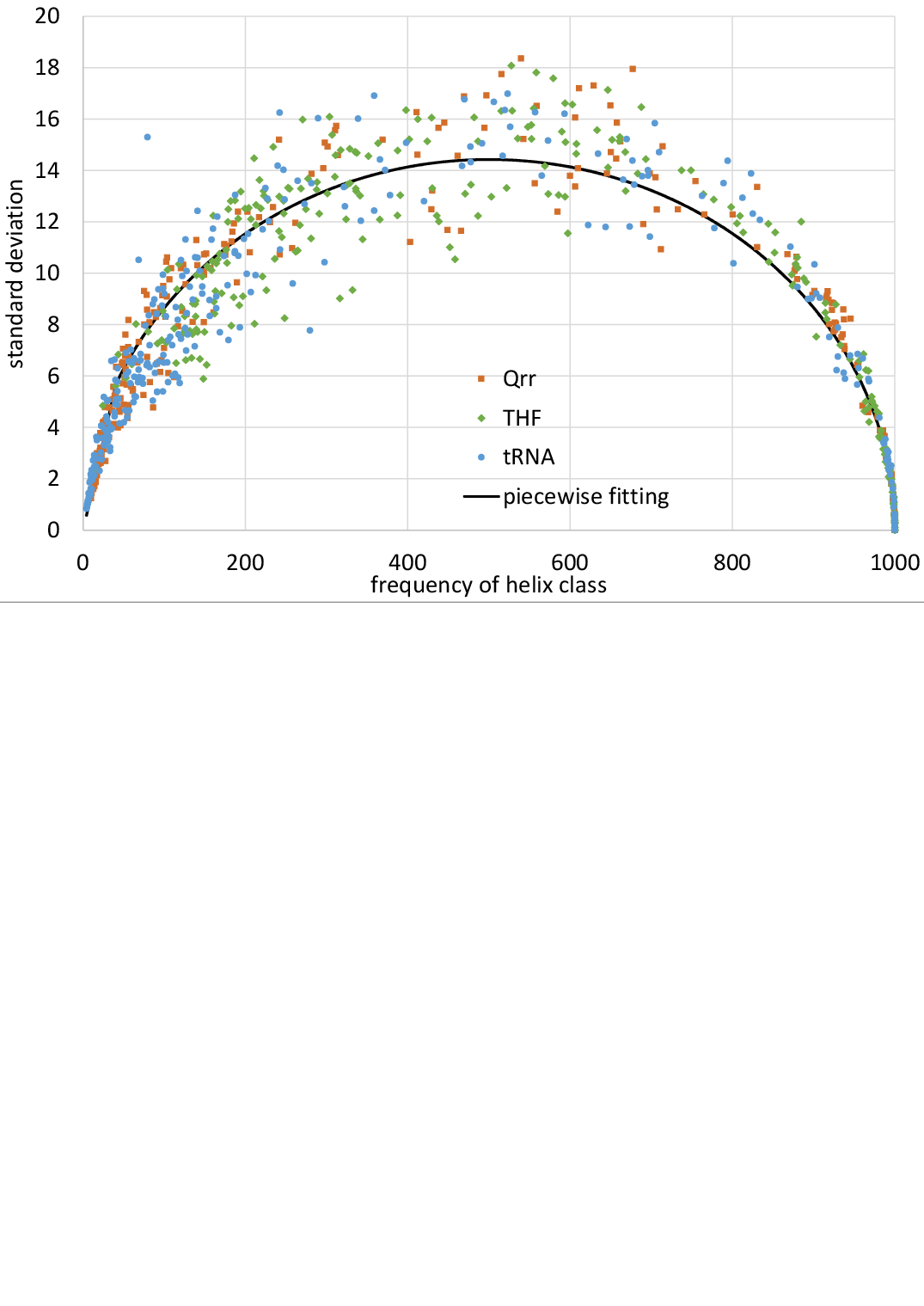} 
    \vspace{-9cm}
    \caption{The correlation between the standard deviation and the average of helix class frequency with 100 replications across 3 types of RNAs and the piecewise fitting curve. 25 Qrrs, 20 THFs, and 22 tRNAs chosen at random are used to find this correlation.}
    \label{piecewise}
\end{figure}

In order to predict with only one Boltzmann sample how the number of features changes, we simulate how the number of features changes upon replication. The number of features is determined by where the maximum cumulative average Shannon entropy occurs, and Shannon entropy is calculated from the helix class frequencies. Therefore, it is necessary to find the distribution of the helix classes frequencies. By taking 100 samples for 67 RNA sequences and recording the frequencies of their first 12 helix classes, we found that there is a correlation between the frequency of a helix class and its standard deviation and this correlation is valid for different classes of RNAs. This correlation is shown in Fig. \ref{piecewise}. This correlation is what we will exploit in order to predict stability.
\begin{figure}
    \centering
    \includegraphics[width=14cm]{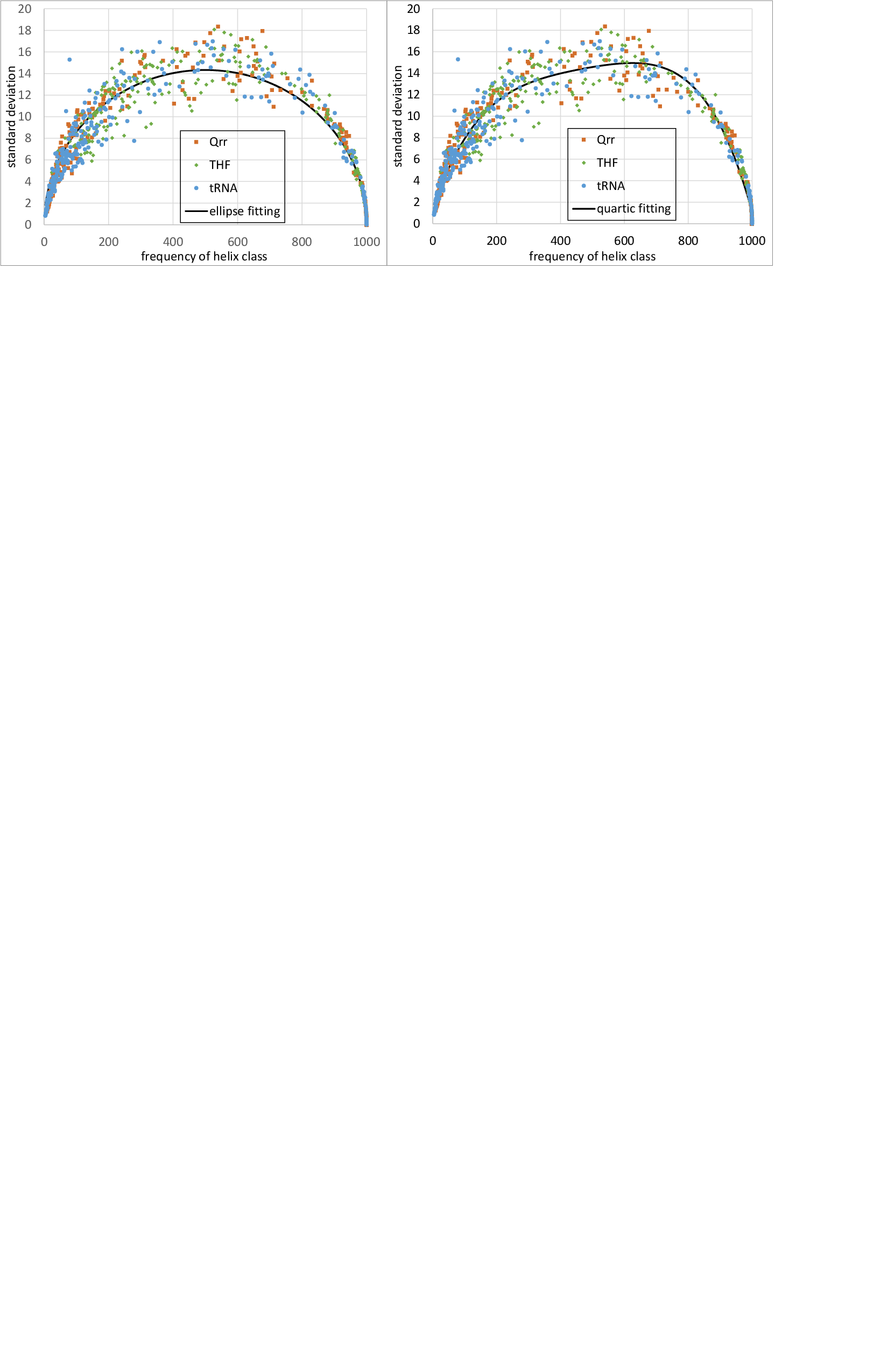} 
    \vspace{-18cm}
    \caption{Ellipse(left) and quartic(right) curves for the standard deviation and the average of helix class frequency with 100 replications.}
    \label{ellipse_quartic}
%     \begin{figure}%
%     \centering
%     \subfloat[label 1]{{\includegraphics[width=5cm]{quartic.png} }}%
%     \qquad
%     \subfloat[label 2]{{\includegraphics[width=5cm]{fitting.png} }}%
%     \caption{2 Figures side by side}%
%     \label{fig:example}%
% \end{figure}
\end{figure}

\section{Methods}

We fit several equations to the data points from Fig. \ref{piecewise}. We tried 3 kinds of equations: quartic, ellipse and a piecewise function made of a linear portion and an ellipse. 
The best fit quartic equation is 
\begin{equation} 
        y=1.4658+0.084934x-2.3496\times10^{-4}x^2+3.206\times10^{-7}x^3-1.7092\times10^{-10}x^4
        .\end{equation} 
The quartic equation gives a good fit of the data, but is not biologically motivated, so we tried to use other curves. It is reasonable to make (0,0) and (1000,0) points on our fitting curve, since if a helix class average frequency is 0 or 1000, then it will never occur or always occur and will have standard deviation of 0. The best fit ellipse equation is 
\begin{equation} 
        y=14.3348\sqrt{1-(\frac{x-500}{500})^2}
        .\end{equation} 

The ellipse equation is more natural, but the data points with x\textless50 are clearly below the curve, so we also used a piecewise function with a straight line across the origin at the beginning and an ellipse curve for the rest.The best fit piecewise function is 
\begin{equation} 
y=\begin{cases} 
		0.140974x & 0\leq x\leq 40.223;\\
      	14.4298\sqrt{1-(\frac{x}{500})^2} & 40.223< x\leq1000 
   \end{cases}
 .\end{equation} 
The parameters for the three equations were determined by least squares fitting.

We developed 3 methods to predict the stability of the number of features using only one replication of RNAStructProfiling. The main idea is to perturb the frequencies and watch whether the number of features changes. We then compared our predicted result with the result we got from 100 replications of RNAStructProfiling.

Assume we have the frequencies of the features from one replication of RNAStructProfiling having $m$ helix classes and $F$ features. The frequencies of helix classes in descending order are put in a list \{$f_1$, $f_2$, $f_3$, \ldots, $f_F$, \ldots, $f_m$\}. We take the first $F+6$ frequencies as our input and try to predict whether it will be stable or not. 

\subsection{Fixed method}
The first method is called the fixed method. In this method, we try to generate a simulated list of frequencies of 100 replications with only one replication. We predict the standard deviation $\sigma_k$ of the frequency, $f_k$, based on the relationship between the standard deviation and the value of the frequencies plotted in Fig. \ref{piecewise}.

We then take a random value from the distribution $\mathcal{N}(f_k,\,\sigma_k^{2})$ to simulate the possible frequency in one replication. This process is done for the first $F+6$ helix classes. Then this process is repeated 100 times to generate 100 predictions of frequency lists. The first two perturbed frequency lists for VvQrr1 (Vibrio vulnificus qrr1\cite{Vvqrr1}) are shown in Table \ref{example}. The number of features can be calculated from the predicted frequencies for each replication using Eqs. (\ref{pk}), (\ref{Hk}), and (\ref{hk}). We can estimate the standard deviation of the number of features. If the standard deviation is 0, then we predict that the result is stable. For VvQrr1, the number of features is 7 for all 100 simulations, so it is predicted to be stable. 

In order to find whether the distribution of frequency of a helix class is normal, we chose 2 RNAs at random from all RNAs used in this paper and tested them with the Anderson-Darling test. The frequencies of the first 13 helix classes from \textit{Clostridiales bacterium} KA00274 THF and \textit{Homo sapiens} mitochondrially encoded tRNA glycine (MT-TG) riboswitch were chosen and tested. With the confidence level of 95\%, 7 out of 26 of the frequencies are not normally distributed(Appendix Table \ref{normality}). We can assume the 19 of the 26 frequencies are normally distributed. Although some helix classes did not pass the test, we assumed that the frequency $f_k$ of the helix at index $k$ across different samples is normally distributed, since it was the best assumption we found.

\begin{table}[]
\centering
\caption{Frequencies from one replication of RNAStructProfiling and the first two simulated of frequency lists for the first 13 helix classes of VvQrr1 using the fixed method. The bold numbers are the cutoff.}
\label{example}
\begin{tabular}{|c|c|c|c|}
\hline
index & sample data & prediction 1 & prediction 2 \\ \hline
1     & 997         & 998          & 997          \\ \hline
2     & 995         & 994          & 995          \\ \hline
3     & 911         & 911          & 919          \\ \hline
4     & 867         & 864          & 866          \\ \hline
5     & 629         & 624          & 636          \\ \hline
6     & 486         & 463          & 472          \\ \hline
7     & \textbf{298}         & \textbf{267}          & \textbf{313}          \\ \hline
8     & 99          & 124          & 111          \\ \hline
9     & 85          & 82           & 77           \\ \hline
10    & 59          & 59           & 57           \\ \hline
11    & 53          & 47           & 52           \\ \hline
12    & 46          & 45           & 44           \\ \hline
13    & 38          & 45           & 33           \\ \hline
\end{tabular}
\end{table}

\subsection{Variable method}
The second method is called the variable method. This method tries to test the the tolerance of the frequency list to possible changes. The preliminary work is similar to the first method. Instead of taking a random value from the distribution $\mathcal{N}(f_k,\,\sigma_k^{2})$, we simulate a frequency from the distribution $\mathcal{N}(f_k,\,a_i^2\sigma_k^{2})$. The multiplier, $a_i$, is taken from a geometric series with first term $a_0=0.1$ and ratio $q=1.2$. The two parameters are chosen to generate a wide-range list with clear classification for RNAs with different stabilities. We generate 10 simulated of frequency lists and watch whether the number of features changes. If it changes, we will record the current $a_i$; otherwise, we will move to $a_{i+1}$ until a change occurs. If the recorded $a_i$ is small, then the sample is sensitive to the variation of the frequencies, which means the number of features is likely to change across replications of RNAStructProfiling. If the recorded $a_i$ is large, then the sample will be stable under variations. 

We tested several values of threshold for the cutoff, $c$, and the correct prediction rates with different cutoffs are shown in Table \ref{3methodsQrr}. $c$ is set at 1.9, so that we can get the maximum correct prediction rate in the test for Qrrs. In the test for VvQrr1, the number of features stays at 7 for all of the 10 replications when $a\leq1.85$. The first change occurs when $a$ increases to 2.22, so the a is 2.22 for VvQrr1. Since $a>c=1.9$, this method predicts that VvQrr1 is stable.

\begin{table}[]
\centering
\caption{Correct prediction rate for Qrrs using quartic equation.}
\label{TFerror}
\begin{tabular}{|c|c|c|}
\hline
                                           & fixed & variable \\ \hline
P(correct prediction$|$predicted stable)   & 0.582        & 0.837        \\ \hline
P(correct prediction$|$predicted unstable) & 0.934        & 0.926        \\ \hline
P(correct prediction)                      & 0.833        & 0.910        \\ \hline
\end{tabular}
\end{table}
% 0.625	0.740
% 0.941	0.929
% 0.854	0.888
% 0.897	
\subsection{Hybrid method}
When we tested the two methods, we found that, as shown in Table \ref{TFerror}, both of them have higher reliability for the cases when the predicted result is unstable and a relatively low correct rate for the cases of stable predictions. The correct rate for stable prediction is higher for variable than for fixed, while the unstable prediction has a higher correct rate for fixed than for variable. Therefore, we developed the third method called hybrid method. The hybrid method checks if the fixed prediction is unstable. If the fixed method predicts it is unstable, then the correct rate is high (in the case of Qrrs, 0.934) and the hybrid method will also predict unstable. Otherwise, it will use the prediction from the variable method. The accuracy of the hybrid method prediction for the case of Qrrs is 0.906. We also tested the hybrid method with different fitting equations and cutoff $c$'s. The correct prediction rates for Qrrs are in Table \ref{3methodsQrr} and the results for THF, tRNA and TPP are in Appendix Table \ref{THF}, \ref{tRNA} and \ref{TPP}.
\begin{table}[]
\centering
\caption{Correct prediction rate for Qrrs using different fitting equations and different cutoff between small and large $a$'s.}
\label{3methodsQrr}
\begin{tabular}{|c|c|c|c|c|}
\hline
\multicolumn{2}{|c|}{}          & piecewise& ellipse & quartic  \\ \hline
\multicolumn{2}{|c|}{fixed}     
								& 0.876   & 0.876   & 0.833     \\ \hline
\multirow{3}{*}{variable}    
						& c=1.9 & 0.863   & 0.901   & 0.910     \\ \cline{2-5} 
                        & c=1.6 & 0.863   & 0.876   & 0.858     \\ \cline{2-5} 
                        & c=1.3 & 0.820   & 0.828   & 0.833     \\ \hline
\multirow{3}{*}{hybrid} 
						& c=1.9 & 0.863   & 0.906   & 0.906     \\ \cline{2-5} 
                        & c=1.6 & 0.876   & 0.906   & 0.880     \\ \cline{2-5} 
                        & c=1.3 & 0.876   & 0.893   & 0.867     \\ \hline
\end{tabular}
\end{table}
% \begin{table}[]
% \centering
% \caption{Correct prediction rate for 4 types of RNAs using fixed method.}
% \begin{tabular}{|c|c|c|c|}
% \hline
%      & piecewise & ellipse & quartic   \\ \hline
% Qrr  & \textbf{0.876}   & 0.876   & 0.833      \\ \hline
% THF  & 0.915   & \textbf{0.923}   & \textbf{0.931}      \\ \hline
% tRNA & \textbf{0.840}   & \textbf{0.839}   & \textbf{0.871}      \\ \hline
% TPP  & 0.910   & \textbf{0.938}   & 0.904       \\ \hline\hline
% average  & 0.885   & 0.894   & 0.885       \\ \hline
% \end{tabular}
% \end{table}

\section{Results} 

\begin{table}[]
\centering
\caption{Correct prediction rate for 4 types of RNAs using 3 methods.}
\label{comparision}
\begin{tabular}{|c|c|c|c|c|}
\hline
\multicolumn{2}{|c|}{}           & piecewise & ellipse & quartic   \\ \hline
\multirow{4}{*}{fixed}    & Qrr  & \textbf{0.876}   & 0.876   & 0.833  \\ \cline{2-5} 
                          & THF  & 0.915   & \textbf{0.923}   & \textbf{0.931}  \\ \cline{2-5} 
                          & tRNA & \textbf{0.840}   & \textbf{0.839}   & \textbf{0.871}  \\ \cline{2-5} 
                          & TPP  & 0.910   & \textbf{0.938}   & 0.904  \\ \hline
\multirow{4}{*}{variable} &Qrr  & 0.863   & 0.901   & \textbf{0.910}      \\ \cline{2-5} 
                          &THF  & 0.915   & 0.900   & \textbf{0.931}      \\ \cline{2-5} 
                          &tRNA & 0.800   & 0.815   & 0.806      \\ \cline{2-5} 
                          &TPP  & \textbf{0.944}   & 0.932   & \textbf{0.944}       \\ \hline
\multirow{4}{*}{hybrid}   & Qrr  & 0.863   & \textbf{0.906}   & 0.906       \\ \cline{2-5} 
                          &THF  & \textbf{0.923}   & 0.900   & \textbf{0.931}      \\ \cline{2-5} 
                          &tRNA & 0.800   & 0.808   & 0.800      \\ \cline{2-5} 
                          &TPP  & \textbf{0.944}   & \textbf{0.938}   & \textbf{0.944}       \\ \hline
\end{tabular}
\end{table}

We tested the prediction methods and fitting curves with 233 Qrr RNAs, 130 THF RNAs, 127 tRNAs, and 177 TPP RNAs. Using 100 replications of RNAStructProfiling, the percentage of stable RNAs for each class is 21.15\%, 11.5\%, 25.5\%, and 16.9\%, respectively. The correct prediction rate for predicting the stability of the 4 types of RNAs are shown in Table \ref{comparision}. The best correct prediction rate for the same type of RNA with the same fitting curve among the 3 methods are in bold. The correct prediction rate is higher for the RNA types with higher percentage of unstable RNAs, because the correct prediction rate for predicting unstable is higher than for predicting unstable.

The hybrid method was intended to combine the strengths of the two methods. It gives the best correct prediction rate half of the time. The fixed method is more reliable for predicting tRNA, which has a high percentage of stable RNAs. The fixed method gives the best correct prediction rate slightly higher than half of the time. The variable method only gives the best correct prediction rate 4 out of 12 times. Therefore, we believe the fixed method is the most reliable and versatile overall. 

Among the 3 fitting curves, we recommend using ellipse. In the fixed method, the average correct prediction rate with piecewise, ellipse and quartic are 0.885, 0.894 and 0.885 respectively. We observe that the average correct prediction rate for ellipse fitting is the highest. The rages of correct prediction rate with piecewise, ellipse and quartic fitting are 0.840-0.915, 0.839-0.938 and 0.833-0.931, respectively. Ellipse fitting also gives the best correct prediction rate generally.

The prediction methods avoid repeated sampling, and thus save considerable computation time. We tested the time for our predicting methods and 100 replications and the results are in Table \ref{time}. On average, the prediction of stability only takes 0.2\% of the time for one profiling process. It can give a high correct prediction rate without 100 replications, which takes about 500 times longer than our prediction method. The time for fixed method is relatively stable across different RNAs. The time for variable method is longer for the stable RNAs, since we stop when a change in $a$ occurs and stable RNAs require us we to try more $a$'s.

\begin{table}[]
\centering
\caption{Time for repeating 100 sampling and profiling process and predicting methods for 6 RNA sequences of varying length.}
\label{time}
\begin{tabular}{|  >{\centering}m{4cm}|>{\centering}m{1.5cm} |>{\centering}m{1.7cm} | c | c |}
\hline
\multicolumn{2}{|c|}{} & \multicolumn{3}{c|}{time} \\ \hline
RNA & length of RNA & repeating 100 times & fixed & variable \\ \hline
Homo sapiens mitochondrially encoded tRNA glycine (MT-TG)&68            & 18.14               & 0.000256 & 0.000244 \\ \hline
Homo sapiens mitochondrially encoded tRNA methionine (MT-TM)&68            & 12.76               & 0.000503 & 0.000551 \\ \hline
Vibrio vulnificus Qrr1&93            & 17.52               & 0.000476 & 0.000930 \\ \hline
Firmicutes bacterium CAG:475 THF riboswitch&94            & 33.27               & 0.000504 & 0.000071 \\ \hline
Ruminococcus sp. CAG:382 THF riboswitch&129           & 44.36               & 0.000430 & 0.000584 \\ \hline
Burkholderia pseudomallei K96243 TPP riboswitch (THI element)&132           & 51.63               & 0.000396 & 0.000391 \\ \hline
\end{tabular}
\end{table}

\section{Acknowledgments and Future Work}

Special thanks to an anonymous referee whose suggestions improved the paper in a fundamental way. Also thanks to Dr. Eric Reyes for suggesting that the cut-off for stability be based on the histogram. 
This research is supported by R-SURF (Rose-Hulman Summer Undergraduate Research Fellowships). We want to thank the research groups from the Georgia Institute of Technology who provided the GTfold and RNAStructProfiling software. We also want to thank Desmond Davis from Rose-Hulman Institute of Technology for his help in compiling and building the environment for the programs.

In future work, one of the authors plans to investigate a related question: Rather than predicting which sequences are stable or unstable under RNAprofiling (which uses a sample of size 1000), we hope to estimate for a given sequence how large a sample needs to be to ensure that a sample is stable.

In other possible future work, we would hope to develop a algorithm to predict the stability of the number of selected profiles, which is a profiling signal at another level. We found that there is not strong relationship between the standard deviation and the value of frequencies of profiles, as shown in Appendix Fig. \ref{profile_sd_vs_freq}, so the algorithm we used for features is not applicable for selected profiles.

\begin{figure}[H]
    \centering
    \includegraphics[width=20cm]{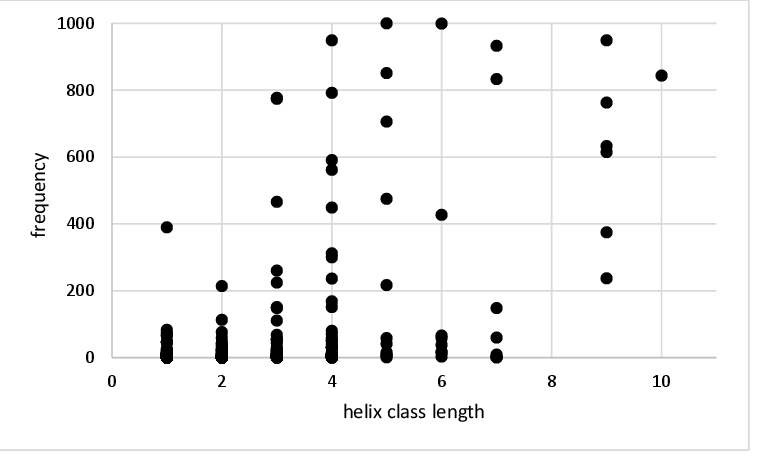} 
    \vspace{-19cm}
    \caption{The relationship between frequency of a helix class and the length of the helix class.}
    \label{freq_vs_length}
\end{figure}

It is an open question to find which structural qualities of the Minimum Free Energy secondary structure are related to the stability of features and profiles. We looked into the relationship between the length of a helix class and its frequency, which was expected to be more positive correlation. The result(Appendix Fig. \ref{freq_vs_length}) shows that there is a positive but not strong correlation between the length of a helix class and its frequency.

We also want to find a biologically motivated equation for the correlation between the frequency and its standard  deviation. Currently, we are using the ellipse equation, but we would also like to find the theoretical backing for why the relation can be fitted with an ellipse equation.

\section{Appendix}
%\newpage

\begin{table}[H]
\centering
\caption{Correct prediction rate for THF riboswitch RNAs using different fitting equations and different cutoff between small and large $a$'s.}
\label{THF}
\begin{tabular}{|c|c|c|c|c|}
\hline
\multicolumn{2}{|c|}{}          & piecewise & ellipse & quartic \\ \hline
\multicolumn{2}{|c|}{fixed}     
								& 0.915   & 0.923   & 0.931     \\ \hline
\multirow{3}{*}{variable}    
						& c=1.9 & 0.915   & 0.900   & 0.931     \\ \cline{2-5} 
                        & c=1.6 & 0.908   & 0.892   & 0.938     \\ \cline{2-5} 
                        & c=1.3 & 0.900   & 0.908   & 0.908     \\ \hline
\multirow{3}{*}{hybrid} 
						& c=1.9 & 0.923   & 0.900   & 0.931     \\ \cline{2-5} 
                        & c=1.6 & 0.923   & 0.900   & 0.946     \\ \cline{2-5} 
                        & c=1.3 & 0.931   & 0.923   & 0.946     \\ \hline
\end{tabular}
\label{THF}
\end{table}
\begin{table}[H]
\centering
\caption{Correct prediction rate for tRNAs using different fitting equations and different cutoff between small and large $a$'s.}
\label{tRNA}
\begin{tabular}{|c|c|c|c|c|}
\hline
\multicolumn{2}{|c|}{}          & piecewise & ellipse & quartic \\ \hline
\multicolumn{2}{|c|}{fixed}     
								& 0.840   & 0.839   & 0.871     \\ \hline
\multirow{3}{*}{variable}
						& c=1.9 & 0.800   & 0.815   & 0.806     \\ \cline{2-5} 
                        & c=1.6 & 0.824   & 0.839   & 0.863     \\ \cline{2-5} 
                        & c=1.3 & 0.856   & 0.839   & 0.847     \\ \hline
\multirow{3}{*}{hybrid} 
						& c=1.9 & 0.800   & 0.808   & 0.800     \\ \cline{2-5} 
                        & c=1.6 & 0.816   & 0.824   & 0.848     \\ \cline{2-5} 
                        & c=1.3 & 0.824   & 0.832   & 0.856     \\ \hline
\end{tabular}
\label{tRNA}
\end{table}
\begin{table}[H]
\centering
\caption{Correct prediction rate for TPP riboswitch RNAs using different fitting equations and different cutoff between small and large $a$'s.}
\label{TPP}
\begin{tabular}{|c|c|c|c|c|}
\hline
\multicolumn{2}{|c|}{}          & piecewise & ellipse & quartic \\ \hline
\multicolumn{2}{|c|}{fixed}     
								& 0.910   & 0.938   & 0.904     \\ \hline
\multirow{3}{*}{variable}    
						& c=1.9 & 0.944   & 0.932   & 0.944     \\ \cline{2-5} 
                        & c=1.6 & 0.949   & 0.955   & 0.932     \\ \cline{2-5} 
                        & c=1.3 & 0.910   & 0.921   & 0.881     \\ \hline
\multirow{3}{*}{hybrid} 
						& c=1.9 & 0.944   & 0.938   & 0.944     \\ \cline{2-5} 
                        & c=1.6 & 0.949   & 0.955   & 0.949     \\ \cline{2-5} 
                        & c=1.3 & 0.938   & 0.949   & 0.927     \\ \hline
\end{tabular}
\label{TPP}
\end{table}

\begin{table}[H]
\centering
%URS0000C03BAE\_1497955 URS00000C6674\_9606
\caption{Anderson-Darling Test result for frequencies of the first 13 helix classes from  \textit{Clostridiales bacterium} KA00274 THF and  \textit{Homo sapiens} mitochondrially encoded tRNA glycine (MT-TG) riboswitch.}
\label{normality}
\begin{tabular}{|c|c|c|c|}
\hline
index & frequency & standard deviation & p-value        \\ \hline
1     & 987.71    & 2.959              & 0.129          \\ \hline
2     & 981.68    & 3.576              & \textbf{0.033} \\ \hline
3     & 879.56    & 10.623             & \textbf{0.007} \\ \hline
4     & 809.22    & 12.236             & 0.608          \\ \hline
5     & 547.99    & 15.701             & 0.949          \\ \hline
6     & 363.29    & 15.066             & \textbf{0.013} \\ \hline
7     & 306.57    & 15.384             & 0.832          \\ \hline
8     & 222.19    & 13.207             & 0.408          \\ \hline
9     & 185.53    & 9.064              & 0.727          \\ \hline
10    & 177.31    & 10.398             & 0.801          \\ \hline
11    & 170.64    & 9.218              & 0.503          \\ \hline
12    & 159.23    & 10.457             & 0.335          \\ \hline
13    & 115.62    & 9.370              & \textbf{0.036} \\ \hline
\end{tabular}

\begin{tabular}{|c|c|c|c|}
\hline
index & frequency & standard deviation & p-value                   \\ \hline
1     & 998.92    & 1.152              & \textbf{\textless{}0.005} \\ \hline
2     & 698.16    & 11.425             & 0.872                     \\ \hline
3     & 673.28    & 11.818             & 0.464                     \\ \hline
4     & 477.05    & 14.924             & 0.538                     \\ \hline
5     & 221.16    & 11.711             & 0.904                     \\ \hline
6     & 187.35    & 13.044             & 0.788                     \\ \hline
7     & 137.97    & 10.620             & 0.254                     \\ \hline
8     & 120.66    & 10.509             & 0.123                     \\ \hline
9     & 38.78     & 4.426              & \textbf{0.018}            \\ \hline
10    & 33.32     & 3.225              & 0.119                     \\ \hline
11    & 29.52     & 3.783              & \textbf{0.036}            \\ \hline
12    & 28.61     & 4.420              & 0.108                     \\ \hline
13    & 26.07     & 3.301              & 0.122                     \\ \hline
\end{tabular}
\end{table}

\begin{figure}[H]    
    \centering
    \includegraphics[width=20cm]{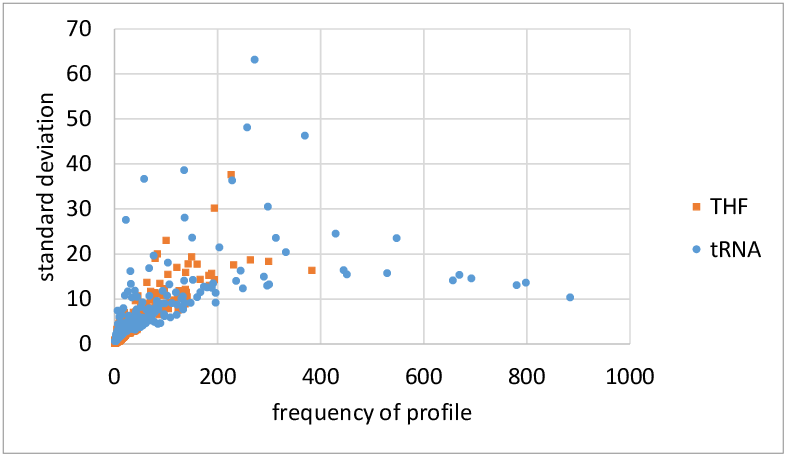} 
    \vspace{-19cm}
    \caption{The relationship between frequency of a profile and its standard deviation with 100 replications for 20 tRNAs and 18 THFs.}
    \label{profile_sd_vs_freq}
\end{figure}

\end{document}